\documentclass[prb,a4paper,10pt,twocolumn,floatfix,superscriptaddress,amsmath,amsfonts,amssymb,preprintnumbers,citeautoscript]{revtex4}
\pdfoutput=1
\usepackage[T1]{fontenc}\usepackage[latin1]{inputenc}
\usepackage{dcolumn,graphicx,color,booktabs,microtype,afterpage} 
\usepackage{sidecap}
\usepackage{times}
\usepackage[colorlinks,plainpages=false,linkcolor=blue,urlcolor=blue,citecolor=blue,pdfpagemode=UseNone,pdfstartview=FitBH]{hyperref}

\usepackage{amssymb}
\usepackage{amsmath}
\usepackage{mathtools}
\usepackage{graphicx}

\newcommand{\dersio}{D-Er$_{2}$Si$_{2}$O$_{7}$}
\newcommand{\ybsio}{Yb$_{2}$Si$_{2}$O$_{7}$}

\newcommand{\invang}{\AA$^{-1}$}

\begin{document}
%
%
%
%
\title
{Magnetic Properties of the Ising-like Rare Earth Pyrosilicate: D-Er$_{2}$Si$_{2}$O$_{7}$}
\author{Gavin Hester}
\email{Gavin.Hester@colostate.edu}
\affiliation{Department of Physics, Colorado State University, 200 W. Lake St., Fort Collins, CO 80523-1875, USA}
\author{T. N. DeLazzer}
\affiliation{Department of Physics, Colorado State University, 200 W. Lake St., Fort Collins, CO 80523-1875, USA}
\author{D. R. Yahne}
\affiliation{Department of Physics, Colorado State University, 200 W. Lake St., Fort Collins, CO 80523-1875, USA}
\author{C. L. Sarkis}
\affiliation{Department of Physics, Colorado State University, 200 W. Lake St., Fort Collins, CO 80523-1875, USA}
\author{H. D. Zhao}
\affiliation{Department of Physics, University of Colorado-Boulder, Boulder, Colorado, 80309, USA}
\author{J. A. Rodriguez Rivera}
\affiliation{NIST Center for Neutron Research, National Institute of Standards and Technology, Gaithersburg, Maryland, 20899-6102, USA}
\affiliation{Department of Materials Science and Engineering, University of Maryland, College Park, Maryland 20740, USA}
\author{S. Calder}
\affiliation{Neutron Scattering Division, Oak Ridge National Laboratory, Oak Ridge, TN 37831, USA}
\author{K. A. Ross}
\email{Kate.Ross@colostate.edu}
\affiliation{Department of Physics, Colorado State University, 200 W. Lake St., Fort Collins, CO 80523-1875, USA}
\affiliation{Quantum Materials Program, CIFAR, Toronto, Ontario M5G 1Z8, Canada}
\date{\today}
%
%
%
%
\begin{abstract}
Ising-like spin-1/2 magnetic materials are of interest for their ready connection to theory, particularly in the context of quantum critical behavior. In this work we report detailed studies of the magnetic properties of a member of the rare earth pyrosilicate family, \dersio, which is known to display a highly anisotropic Ising-like g-tensor and effective spin-1/2 magnetic moments.  We used powder neutron diffraction, powder inelastic neutron spectroscopy (INS), and single crystal AC susceptibility to characterize its magnetic properties. Neutron diffraction enabled us to determine the magnetic structure below the known  transition temperature ($T_N = 1.9$ K) in zero field, confirming that the magnetic state is a four-sublattice antiferromagnetic structure with two non-collinear Ising axes, as was previously hypothesized. Our powder INS data revealed a gapped excitation at zero field, consistent with anisotropic (possibly Ising) exchange.  An applied field of 1 T produces a mode softening, which is consistent with a field-induced second order phase transition.  To assess the relevance of \dersio\ to the transverse field Ising model, we performed AC susceptibility measurements on a single crystal with the magnetic field oriented in the direction transverse to the Ising axes.  This revealed a transition at 2.65 T at 0.1~K, a field significantly higher than the mode-softening field observed by powder INS, showing that the field-induced phase transitions are highly field-direction dependent as expected. These measurements suggest that \dersio\  may be a candidate for further exploration related to the transverse field Ising model.
\end{abstract}

\maketitle
%
%
%
%
\section{Introduction}
\indent The identification of Ising-like magnetic materials has been historically important for verifying the many intriguing features of the classical Ising model \cite{Wolf2000,fisher1981simple}. In the context of quantum magnetism, examples of Ising materials which can be tuned to a quantum phase transition (QPT) via a transverse magnetic field are in even higher demand. The Transverse Field Ising Model (TFIM) is one of the most tractable models with a QPT, and thus has been studied extensively theoretically, including in the burgeoning field of non-equilibrium quantum dynamics \cite{Dutta2015}. However, despite the seemingly straightforward ingredients, there are only a few currently known magnetic materials which approximate the TFIM; CoNb$_{2}$O$_{6}$ (quasi-1D) \cite{Kinross2014}, (Ba/Sr)Co$_{2}$V$_{2}$O$_{8}$ (quasi-1D) \cite{Wang2018, Wang2019}, and LiHoF$_{4}$ (dipolar coupled 3D) \cite{Beauvillain1978, Gingras_2011}. With each of these materials, many detailed comparisons to theoretical expectations have been pursued, and even their non-equilibrium behavior are now being explored \cite{silevitch2019tuning}. Yet, each material has its own deviations from the ideal models, and the identification of additional TFIM materials, particularly those representing the higher dimensional 2D or 3D (non-dipolar) models, which cannot be solved exactly, are of great interest. The first step in finding new TFIM materials is to find materials with predominantly Ising-like exchange. This type of anisotropic exchange is expected to be more prevalent in 4$f$ rare earth based magnetic systems, as the high spin-orbit coupling provides inherent anisotropy to the system.

Indeed, rare earth based materials have become the subject of increased study in the realm of quantum magnetism in general. Due to the chemical similarity of the rare earths, the same structure can often be stabilized with a variety of rare earth ions. However, the magnetic interactions and anisotropies can be dramatically different between each instance; such is the case for the rare earth pyrochlores \cite{Hallas2017, Gardner2010} and the rare earth delafossites \cite{Liu2018, Baenitz2018, Bordelon2019,Xing2019}. In this work we have investigated a member of the rare earth pyrosilicate (RE$_{2}$Si$_{2}$O$_{7}$) family of compounds which have become the subject of renewed interested due to the discovery of a dimer magnet with evidence for a field-induced Bose-Einstein condensate in \ybsio\ \cite{Hester2019, Flynn2020}. The present work focuses on the magnetic properties of one polymorph of Er$_{2}$Si$_{2}$O$_{7}$. It is worth noting that all of the lanthanide series can be synthesized in this stoichiometry (albeit with many possible structures) making this series an interesting playground for understanding the interplay of magnetic species and crystal structure on the ground state of quantum magnets.

 \begin{figure}[h]
 \includegraphics[scale = 0.9]{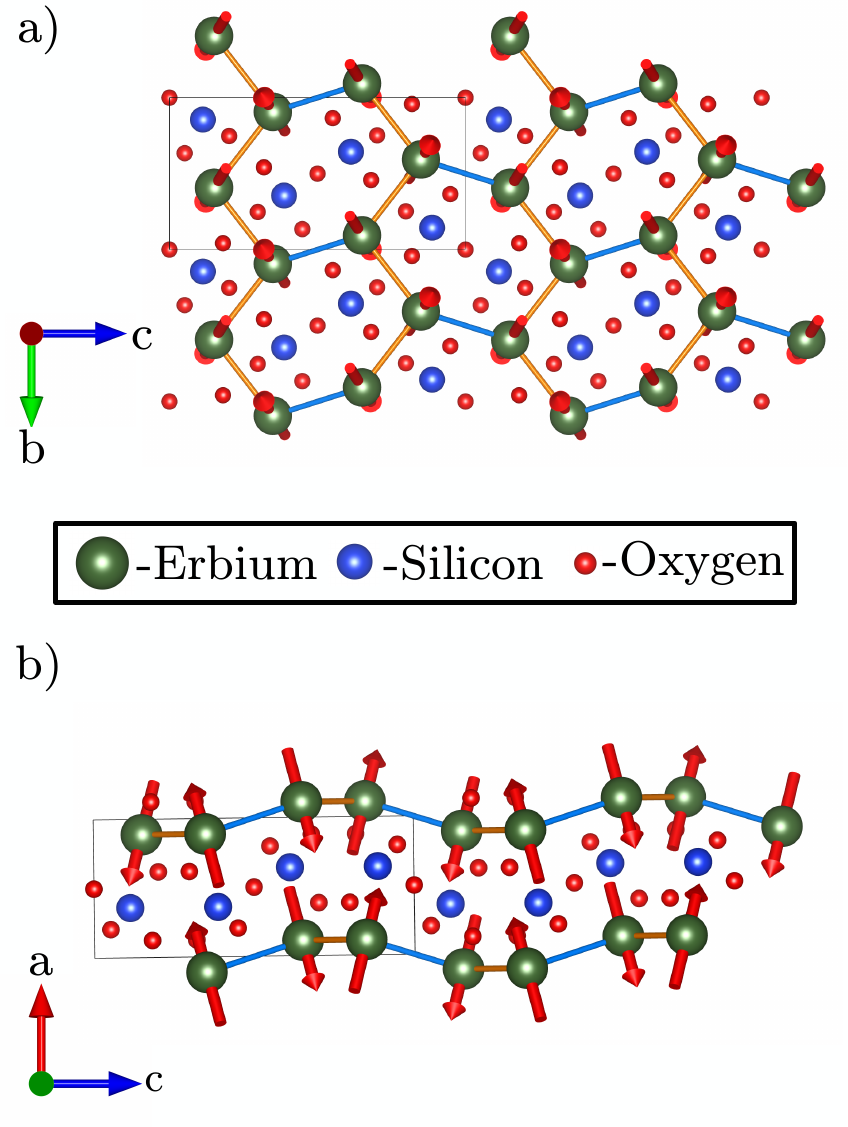}
 \caption{a) The crystal and magnetic structure (obtained from the refinement in Fig.~\ref{fig_elastic}a) of \dersio\ viewed along the $a$-axis. Bonds in orange dictate the equivalent "intrachain" interactions that form chains and bonds in blue dictate the interchain interactions that form a tessellated, distorted honeycomb lattice. Here Er atoms are green, Si are blue, and O are red. b) View of the crystal and magnetic structure along the $b$-axis showing the layered nature of the magnetic ions. All panels of this figure were created using the Vesta software \cite{vesta}.} \label{fig_crystals_dersio}
 \end{figure}

The pyrosilicate compound Er$_{2}$Si$_{2}$O$_{7}$ can crystallize in three different structures based on the synthesis temperature: the low-temperature phase P1 (Type B), the intermediate-temperature phase C2/m (Type C) and the high-temperature phase P2$_{1}$/a (Type D) \cite{Felsche1970, smolin1970crystal}. The focus of this work is the high temperature phase (shown in Fig.~\ref{fig_crystals_dersio}), hereby referred to as \dersio. Since Er$^{3+}$ is a Kramer's ion the ground state of the CEF is protected by time-reversal symmetry and thus must be at least doubly degenerate. In the 1970's the structure of the rare earth pyrosilicates was determined due to interest in the magnetic, electrical, and optical properties of rare earth materials. In particular, the rare earth pyrosilicates were of interest due to the 180$^\circ$ Si-O-Si bond in the [Si$_2$O$_7$]$^{6-}$ groups \cite{Felsche1970,smolin1970crystal}. After some time the magnetic properties of \dersio\ were explored via Zeeman spectroscopy and magnetometry \cite{Maqsood1981,Leask1986}. The Zeeman spectroscopy measurements revealed crystal field levels at 27 cm$^{-1}$ (39 K, 3.3 meV) and 52 cm$^{-1}$ (74 K, 6.4 meV), the former of which we have confirmed via specific heat in Appendix \ref{Cp_SI}. Further, Leask \textit{et al.} determined an Ising-like $g$-tensor anisotropy, with $g_x = 2.6$, $g_y = 3.4$, and $g_z = 13.4$ \cite{Leask1986}. The crystal symmetry and the trivial Er$^{3+}$ site symmetry ($C_{1}$) results in two orientations of these local axes, with the $x$ axis shared by both. The $z$ axis was found to be 28$^\circ$ (clockwise) from the $a$ axis and $\pm$ 15$^\circ$ from the $a$-$b$ plane, while $x$ is in the $a$-$b$ plane (see Fig.~\ref{fig_magnetometry_dersio}c). However, there is a discrepancy in the $g$-tensor values identified by Leask, \textit{et al.} \cite{Leask1986} and those identified earlier by Maqsood \cite{Maqsood1981}. This discrepancy could be due to Maqsood using Curie-Weiss fits to determine the values of the $g$-tensor. Curie-Weiss fits can prove unreliable for rare earth ions due to crystal field effects; typically, they are performed at high temperatures to ensure the system is no longer strongly correlated, but for rare earth ions this causes thermal population of higher crystal field levels and means the fit is not truly indicative of the low temperature angular momentum degrees of freedom. Previous temperature-dependent susceptibility measurements on single crystals of \dersio\ were performed along the $c$ axis, $a$ axis, and the $m$ axis, where $m$ refers to a vector in the $a$-$b$ plane that lies 28$^{\circ}$ (clockwise) from the $a$ axis, which we will refer to as the ``average Ising direction'' (the projection of Leask's $z$ axis onto the $a$-$b$ plane). The susceptibility along all three directions showed a sharp downturn indicative of antiferromagnetic ordering at 1.9 K, with the maximum susceptibility observed for measurements along $m$. We have previously corroborated this magnetic ordering temperature using zero-field specific heat measurements \cite{Nair2019}. Magnetization versus magnetic field measurements along $a$ and $m$ showed evidence of a spin-flip transition at $\frac{1}{3}$ the saturation magnetization. This occurs at $\sim$ 0.5 T for the $a$ axis and slightly lower for the $m$ axis. The observation of a spin-flip transition is consistent with the Ising-like moment found from the $g$-tensor. After these seminal studies of \dersio, no magnetic measurements were performed until the present study.

\section{Experimental Methods}
\begin{figure*}[!t]
\includegraphics[width=2\columnwidth]{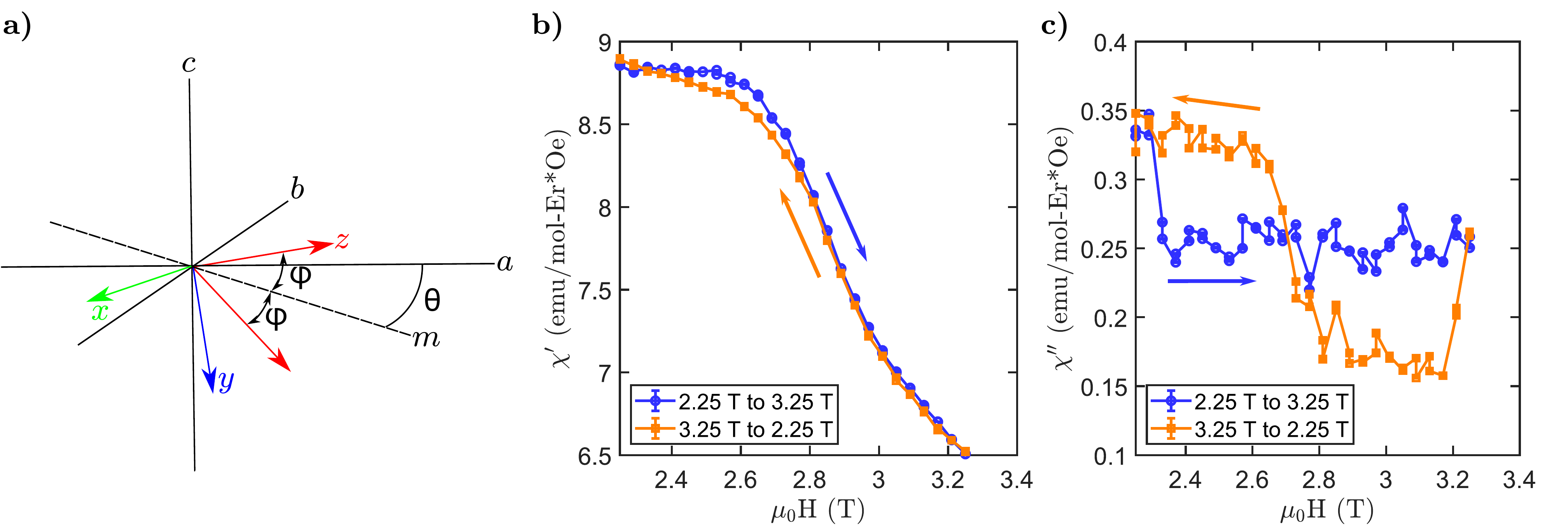}
\caption{a) Schematic diagram showing the average Ising direction ``$m$'' and the two local Ising axes. The $x$-axis is transverse to both Ising axes. Only one of the two $y$-axes is shown for clarity. Here $\theta$ is the angle from the crystallographic $a$-axis and $\phi$ is the angle from the $a$-$b$ plane. Leask \textit{et al.} \cite{Leask1986} previously found the single-ion g-tensor to be oriented  towards $\theta$ = 28$^{\circ}$ and $\phi$ = 15$^{\circ}$. In contrast, we found based on neutron diffraction data that the \textit{ordered} the moments lie at $\theta$ = 21.3$^{\circ}$ and $\phi$ = 12.8$^{\circ}$. For our transverse-field measurements, we chose to align the crystal perpendicular to the \textit{ordered} moment. b-c) The real ($\chi^{\prime}$) and imaginary ($\chi^{\prime\prime}$) components of the AC susceptibility at $T=0.1$ K, $f=1000$ Hz, and $h_{ac} = 2$mT for a field ramp from 2.25 T to 3.25 T (blue circle) and a field ramp from 3.25 T to 2.25 T (orange square). A transition is observed at $\sim$2.65 T, which shows "hysteresis" in both the real and imaginary components of the susceptibility. It is unclear if this hysteresis is an experimental artifact and this is discussed further in the main text. } \label{fig_magnetometry_dersio}
\end{figure*}

Details of the \dersio\ sample synthesis have been outlined elsewhere \cite{Nair2019}, but broadly the synthesis was performed by mixing stoichiometric amounts of Er$_{2}$O$_{3}$ and SiO$_{2}$, pressing the powder into dense rods, and heating the rods between 1400$^\circ$C and 1500$^\circ$C four times with intermediate re-grinding. Two types of samples were used in the present study. For susceptibility measurements, a small single crystal of pure \dersio\ - grown via the optical floating zone technique - was used. The second sample was a powder, which was used for neutron scattering and heat capacity. Rietveld refinement of room-temperature powder x-ray diffraction data indicated that Er$_{2}$SiO$_{5}$, a common (and stubborn) impurity in the synthesis of \dersio, made up approximately 9\% of the sample. The powder x-ray diffraction data for \dersio\ yielded the lattice parameters: $a$ = 4.68878(8) \AA, $b$ = 5.56029(7) \AA, $c$ = 10.79659(10) \AA, $\alpha$ = 90$^{\circ}$, $\beta$ = 90$^{\circ}$, $\gamma$ = 96.043(1)$^{\circ}$. These parameters are consistent with previously published values \cite{smolin1970crystal}. Heat capacity measurements were performed using a Quantum Design PPMS with the heat capacity option on a sintered powder sample of \dersio. AC susceptibility measurements were performed on a single crystal of \dersio\ (as confirmed by Laue x-ray diffraction) using a Quantum Design PPMS with the dilution refrigerator and AC susceptometer. These measurements were performed at numerous temperatures with $f$ = 1000 Hz, H$_{AC}$ = 0.2 mT, and with the DC magnetic field applied transverse to the "average Ising direction" ($x$ in Fig.~\ref{fig_magnetometry_dersio}).

Inelastic neutron scattering (INS) measurements were performed on approximately 5 grams of a sintered powder rod loaded in a Cu canister at the NIST Center for Neutron Research in Gaithersburg, MD, USA using the Multi-Axis Crystal Spectrometer (MACS) \cite{MACS} with a fixed final energy (E$_{f}$) of 2.5 meV, the double focusing monochromator, and Be filters on the incident and scattered beams. Neutron diffraction measurements were performed on the same sample at the High Flux Isotope Reactor at Oak Ridge National Laboratory using the HB-2A (POWDER) diffractometer \cite{hb2a}. The HB-2A data was collected at 10 K, 2 K, and 0.28 K with the Ge(113) monochromator ($\lambda = 2.41$ \AA) and a collimation of open-21'-12'. All errorbars shown in this work indicate $\pm$ one standard deviation.

\section{Results and Discussion}

\subsection{AC Susceptibility}
\label{ac_suscep}
AC susceptibility measurement results (at T = 0.1~K) with a field in the transverse direction (i.e. with AC and DC fields applied along $x$) are shown in Fig.~\ref{fig_magnetometry_dersio}b-c). The directions of the Ising axes ($z$) and transverse direction ($x$) in relation to the crystallographic axes is shown in Fig.~\ref{fig_magnetometry_dersio}a. We note that the magnetization along this transverse field direction of \dersio\ has not been previously studied. For the AC susceptibility measurements, we aligned the sample so that the field was applied along $x$, defined by the angles $\theta$ = 21.3$^{\circ}$ (the angle from the $a$-axis) and $\phi$ = 12.8$^{\circ}$ (the angle from the $a$-$b$ plane) as determined by our neutron diffraction measurements discussed in section~\ref{diffraction_section}. The decision to use the moment direction found via neutron diffraction for the Ising direction - as opposed to the direction found by Leask \textit{et. al} \cite{Leask1986} - was based on the expectation that the ordered moment direction is determined in part by the \textit{exchange} tensor, rather than solely the single-ion g-tensor. A common misconception about the TFIM is that it requires Ising-like single ion anisotropy (e.g. g$_{zz}$ >> g$_{xx}$, g$_{yy}$), but in reality it is only required that the exchange anisotropy be Ising-like.  In a low symmetry local environment like the Er$^{3+}$ site in \dersio, the two Ising axes do not have to be in the same direction.    Hence, we feel that the ordered moment direction is a more appropriate representation of the relevant Ising axis.   

The real ($\chi^{\prime}$) and imaginary ($\chi^{\prime\prime}$) components of the transverse-field susceptibility both show a transition at ~2.65 T. A comparison of the measurements for increasing and decreasing field shows something similar to hysteresis, which would indicate a first order transition. However, the hysteresis does not behave as expected for first-order phase transitions - namely, the transition appears to occur at a \textit{lower} field when the field is increased compared to when it is decreased, which is not expected for first order transitions, which are based on a nucleation and growth mechanism.   Thus, it is currently unclear as to if this hysteresis is an experimental artifact or if it is intrinsic to the sample.  Additional measurements at different temperatures and for wider magnetic field ranges are shown in Section~\ref{AC_SI}.

In a previous preprint version of this manuscript \cite{dersio_arxiv}, a different set of AC susceptibility measurements was presented, which we later determined were obtained with the field not correctly oriented to the transverse ($x$) axis.  This was shown via a subsequent set of DC magnetization measurements on the same crystal which showed a higher saturation magnetization than expected for that direction.  This misorientation was confirmed by Laue x-ray diffraction to be 37$^{\circ}$ away from the transverse axis, approximately lying along the (4$\overline{9}$1) direction.  In the event that these measurements may still be useful to others, we have provided these DC magnetometry and AC susceptibility measurements in the Appendix (Section~\ref{AC_SI}).

\begin{figure*}[htp]
\includegraphics[width=2\columnwidth]{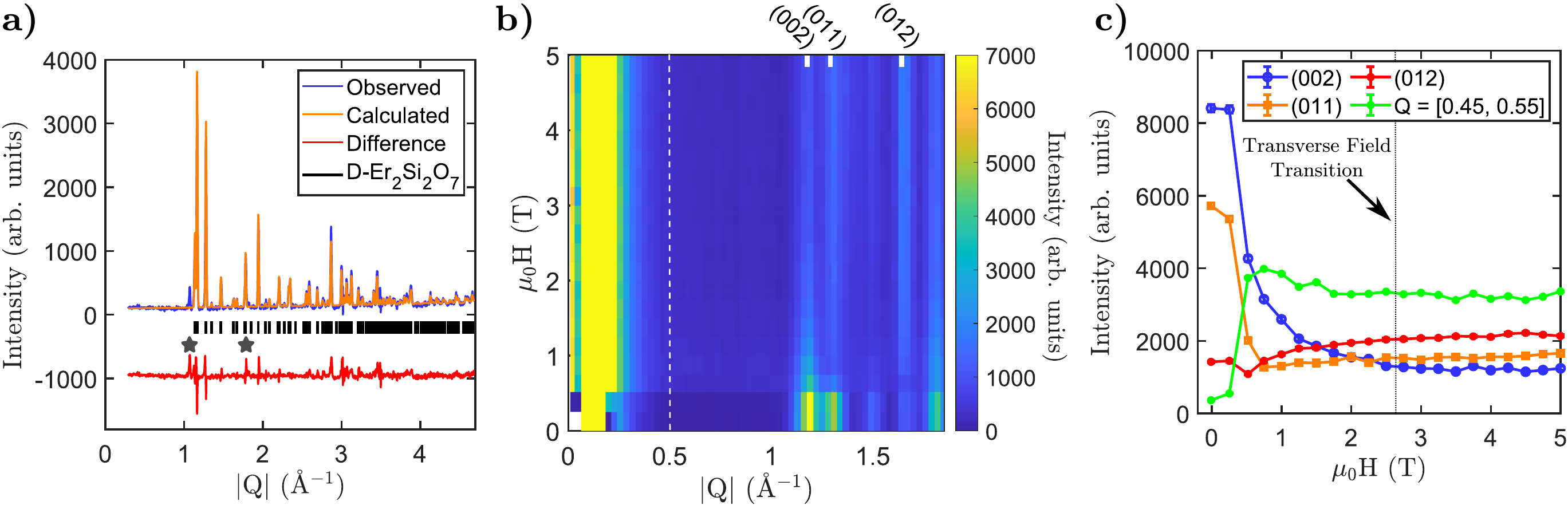}
\caption{a) A Rietveld refinement of the magnetic structure of \dersio. The observed intensities shown are after subtraction of 10 K data from 0.280 K data, and therefore are only due to magnetic order. The data show has an overall shift in the intensity to prevent negative intensity values after subtraction. Two impurity peaks are observed at 1.074 \AA $^{-1}$ and at 1.819 \AA $^{-1}$ - denoted by black stars - that correspond to the (110) and (310) reflections of Er$_{2}$SiO$_{5}$ (space group C2/c). These peaks are the only peaks not accounted for by the magnetic structure of \dersio. b) Field versus momentum transfer ($Q$) elastic scattering (E $\in$ [-0.05,0.05] meV) data obtained on MACS (T$_{avg}$ = 0.16 K) showing the evolution of Bragg peak intensities. White rectangles are shown to denote the Bragg reflections used for cuts in panel c. The white dashed line shows the location of the fourth cut in panel c. c) Evolution of the elastic intensity with field for (002), (011), (012), and Q = [0,45, 0.55] \AA$^{-1}$. The Q = [0.45, 0.55] \AA$^{-1}$ data is scaled by a factor of 10 for clarity. Data at 0 T, 3 T, and 5 T were obtained on the initial increase of the field after cooling from high temperature. All other field points were collected after the field had been increased to 8 T and returned to 0 T. \label{fig_elastic}}
\end{figure*}
\subsection{Neutron Powder Diffraction}
\label{diffraction_section}
Neutron powder diffraction (NPD) data obtained at HB-2A was refined using the FullProf software \cite{fullprof} and the SARAh suite (using the Kovalev tables) \cite{sarah1, sarah2}. Peaks corresponding to \dersio\ and Cu (from the sample can) were observed at 10 K and 2 K (both are above $T_N$), but no sign of the impurity (Er$_{2}$SiO$_{5}$) was observed at these temperatures. This is likely due to the strongest nuclear peaks of Er$_{2}$SiO$_{5}$ occurring at positions obscured by nuclear peaks of \dersio. A Rietveld analysis (Appendix \ref{MagStruc_SI}) of powder neutron diffraction data obtained at 10 K was also performed. The powder diffraction data at 2 K (Appendix \ref{MagStruc_SI}) shows diffuse scattering, as expected for a system approaching a continuous phase transition. Data at 0.280 K (below $T_N$) show an increase in intensity on peaks corresponding to the \dersio\ nuclear structure, aside from two peaks at 1.074 \AA$^{-1}$ and 1.819 \AA $^{-1}$ which can be indexed to the (110) and (310) positions of the impurity phase, Er$_{2}$SiO$_{5}$ (space group C2/c, $a$ = 14.366(2) \AA, $b$ = 6.6976(6) \AA, $c$ = 10.3633(16) \AA, $\alpha$ = 90$^{\circ}$, $\beta$ = 122.219(10)$^{\circ}$, $\gamma$ = 90$^{\circ}$)\cite{phanon2008crystal}. The magnetism of Er$_{2}$SiO$_{5}$ has not previously been reported. Thus, we note in passing that since no magnetic impurity peaks were observed at 10 K or 2 K, the magnetic transition in Er$_{2}$SiO$_{5}$ is likely between 0.280 K and 2 K, to a $|\vec{k}|=0$ ordered state. 
Fig.~\ref{fig_elastic}a shows the results of a Rietveld refinement on the magnetic structure of \dersio. The data used for the refinement was a subtraction of the 10 K data from the 0.280 K data. A symmetry analysis of the allowed $\vert\vec{k}\vert$ = 0 magnetic structures provides four irreducible representations (IR). An attempt to fit each IR was made, with the $\Gamma_{4}$ representation providing the best fit. The $\Gamma_{4}$ IR consists of three basis vectors; $\psi_{10}$, $\psi_{11}$, and $\psi_{12}$ (see Section~\ref{MagStruc_SI} for the basis vector compositions). The coefficients for each basis vector are; C$_{10}$ = 5.72(3), C$_{11}$ = -2.34(6), and C$_{12}$ = -1.45(6). These coefficients yield the magnetic structure shown in Fig.~\ref{fig_crystals_dersio}. This structure has the magnetic space group of P2$_{1}$'/c. Leask \textit{et. al.} found the $g$-tensor axes to lie along $\theta$ = 28$^{\circ}$ (the angle from the $a$-axis) and $\phi$ = 15$^{\circ}$ (the angle from the $a$-$b$ plane), whereas, we have found the moments to lie at $\theta$ = 21.3$^{\circ}$ and $\phi$ = 12.8$^{\circ}$. This indicates that the ordered moments deduced from our refinement lie 6.9(5)$^{\circ}$ away from the single-ion Ising direction determined by Leask \textit{et. al.} However, the overall ordered moment is similar. We measured an ordered moment of 6.56(3) $\mu_{B}$ at 250 mK and the moment found by Leask was 6.7 $\mu_{B}$ at 4.2 K. The discrepancy in the Ising axis direction is potentially due to the presence of the (unaccounted for, but small) Er$_{2}$SiO$_{5}$ impurity in our magnetic NPD data. Alternatively, it could be due to a difference in the exchange anisotropy directions compared to the single-ion anisotropy directions.  The direction of the ordered moment direction would be influenced by both types of anisotropies.  The details of the interactions in \dersio\ require further study, ideally by INS on single crystal samples.

\subsection{Field-dependent Elastic Neutron Scattering}

Elastic neutron scattering data measured using MACS, using the same powder sample as used at HB-2A, is shown in Fig.~\ref{fig_elastic}b and Fig.~\ref{fig_elastic}c. An ``empty can'' background was subtracted. Fig.~\ref{fig_elastic}b shows the field evolution of the elastic scattering at at an average temperature (T$_{avg}$) = 0.16 K. Magnetic Bragg peaks from the impurity (Er$_{2}$SiO$_{5}$) are not resolved, due to the coarser $Q$-resolution of MACS compared to HB-2A. Intensity versus field cuts for the (002), (011), and (012) reflections are shown in Fig.~\ref{fig_elastic}c. The intensity of the peaks does not change significantly between 0~T and 0.25~T, even though the 0.25~T data was measured after going to high field (see caption to Fig.~\ref{fig_elastic} for more detail). Dramatic changes in magnetic peak intensities are observed between 0.25~T and 2~T. Since the experiment was performed on a powder sample, all field directions are averaged here, including the field perpendicular to the Ising direction, which we have shown induces a transition at 2.65 T (Fig.~\ref{fig_magnetometry_dersio}). Other field directions are already known to induce transitions in the field range of about 0.5~T \cite{Leask1986}. Overall, the dramatic changes in Bragg intensity all occur below about 1~T. Above 1~T, the intensity of the peaks gradually changes, likely due the Er$^{3+}$ moments approaching saturation along the field direction, as much as would be allowed by the Ising axes. An interesting effect is observed away from the Bragg peaks, shown in Fig.~\ref{fig_elastic}c as a cut at Q = 0.5 \invang (integrated from 0.45 to 0.55 \invang). The incoherent background appears to increase in intensity at 0.5~T and beyond. The exact nature of this signal is not currently known, but it is not due to an irreversible change in sample environment such as water condensation on the cryostat (we have confirmed this by comparing to scans at similar field ranges taken before and after the dataset shown in Fig.~\ref{fig_elastic}c). It is possible that this background is indicative of short-range spin correlations. Near a second order transition, one does expect the critical behavior to manifest as a sharp increase of diffuse scattering. However, this diffuse scattering should diminish rapidly away from the transition, which we do not observe on the high-field side (the intensity seems to plateau after 2 T). We do not see any signatures of the transverse field transition at $\sim$2.65 T in our elastic neutron scattering data. It is likely that the transition field strongly depends on the field direction, thus very few grains in the powder would be in the correct orientation to probe this transition, leading to a small signal in the neutron scattering measurements.

\begin{figure*}[htp]
\includegraphics[width=2\columnwidth]{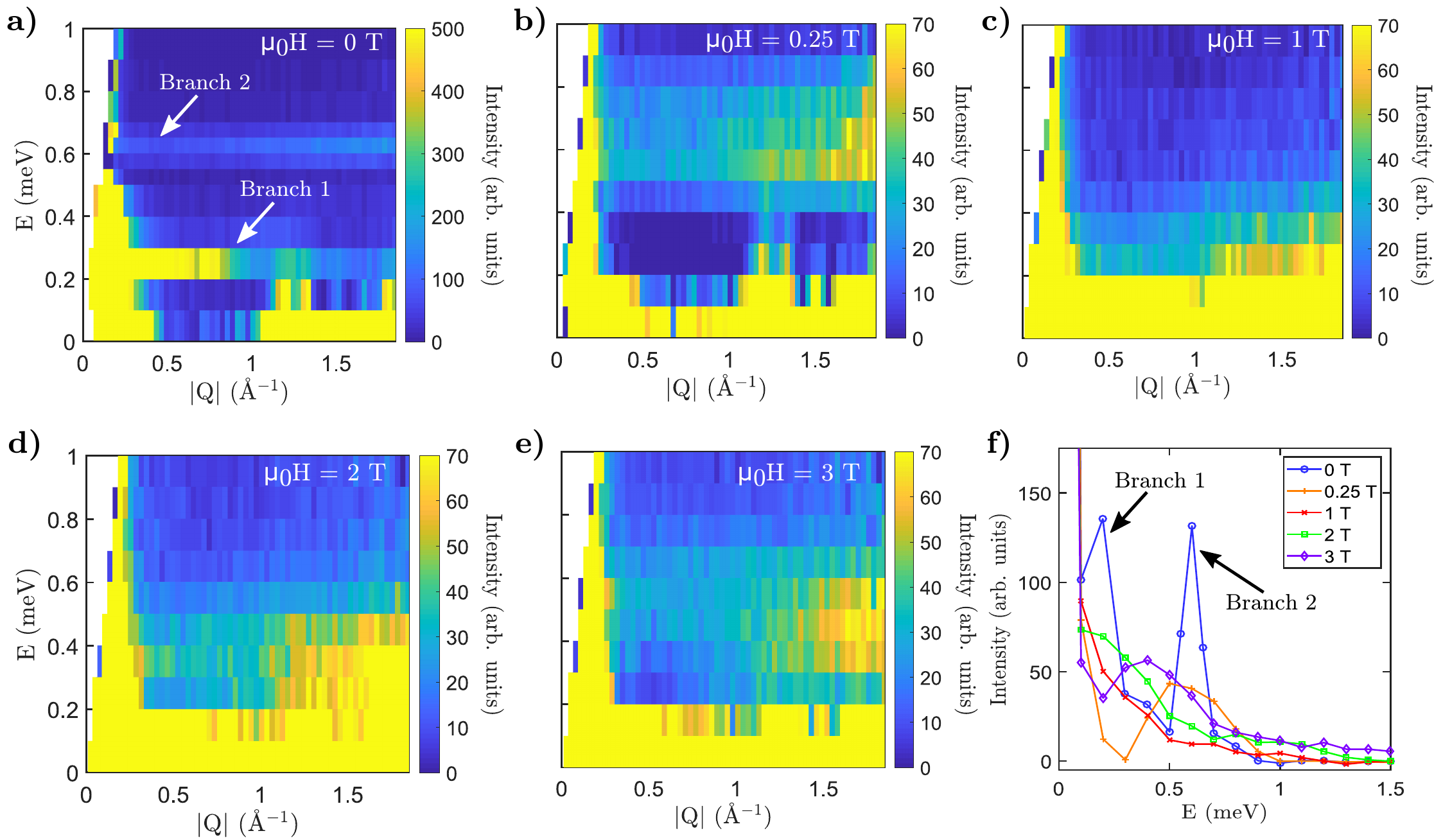}
\caption{(a-e) Energy vs $\vert Q \vert$ slices at $\mu_{0}$H = 0 T (a), 0.25 T (b), 1 T (c), 2 T (d), and 3 T (e) showing the evolution of two excitations. The average temperature for these slices are T$_{avg}$ = 0.22 K, 0.18 K, 0.16 K, 0.17 K, and 0.18 K, respectively. Note: the scale factor for panel a) is higher than the other panels in order to make Branch 1 visible. Branch 1 is only visible in our 0 T data, while Branch 2 is observed to broaden and soften as the field increases to 1 T. f) Intensity vs energy cuts at $Q$ = [1.45,1.55] \AA$^{-1}$\, at various field strengths.} \label{fig_inelastic}
\end{figure*}

\subsection{Inelastic Neutron Scattering}

Finally we turn to the INS data from MACS, shown in Fig.~\ref{fig_inelastic}a-f. In zero field, two branches of excitations are observed at 0.2 meV and 0.6 meV, which we refer to as Branch 1 and Branch 2, respectively. The branches are gapped and appear to be dispersionless within the resolution of the measurement (the energy step size was chosen to be 0.1 meV, which is similar to the instrument resolution for our settings, $\sim$0.075 meV \cite{MACS}). The gapped nature of the excitations is consistent with anisotropic interactions, such as Ising exchange, and the presence of two branches implies at least two non-equivalent sites in the magnetic unit cell. This can be easily understood based on the two local Ising axes. As discussed above, we have found a $\vert\vec{k}\vert$ = 0 antiferromagnetic structure that consists of AFM moments which are collinear when considering sites with the same Ising axis orientation. Thus, each branch is expected to be two-fold degenerate. Something that is more difficult to understand is that as the field increases, we observe only one branch clearly, despite there still being two distinct Ising axes. At 0.25 T (Fig.~\ref{fig_inelastic}b), only Branch 2 remains, and is significantly broadened. The disappearance of Branch 1 might indicate a very low field transition for some field directions, but Leask \textit{et. al} did not report any magnetic transitions below 0.5 T at 500 mK. However, Leask \textit{et. al} did predict that the transition near 0.5 T was first order in nature for some field directions (significant hysteresis was observed in the simulated magnetization). The absence of Branch 1 in our 0.25 T data may be related to this predicted first order transition; the 0.25 T data was collected after the sample was subjected to a very high field (8 T), thus the data at 0.25 T represents the decreasing field part of the hysteresis curve, which is likely to be in a different state than the 0 T data. However, to understand all the details of the field evolution of excitations would require further study on single crystal samples.

We concern ourselves now with the behavior of Branch 2 at higher fields, which exhibits clearer signatures. As the field is increased to 1 T, Branch 2 broadens (likely due to the anisotropic $g$-tensor) and softens dramatically. The excitation becomes gapless near 1 T as expected for a field-induced second order phase transition. Beyond $\sim$ 1 T, the branch energy increases again, consistent with entering the field-polarized paramagnetic regime. No signature of the transverse field transition is observed in the inelastic spectrum. It is worth noting that we have not attributed any signatures in the INS spectrum to the impurity (Er$_{2}$SiO$_{5}$), even though it is also magnetic. Due to the relatively low concentration in the sample (9\%), we expect the impurity will not contribute appreciably to the INS signal.

We also obtained INS data at fields up to 8 T (see Appendix \ref{MACS_SI}), where the excitations could in principle be modelled by linear spin wave theory. Additionally, Leask \textit{et. al.} obtained exchange interactions based on a mean field approach to describing features in the magnetization curves (these parameters, when used in a Monte Carlo simulation, did reproduce many of the observed features of the magnetization curves in several directions \cite{Leask1986}). Unfortunately, the exchange interactions as reported in that work are not uniquely assignable to specific pairs of ions in the unit cell, so a useful comparison to our data is greatly complicated. We would like to note that at least the interlayer interaction (along $a$) seems to be clearly defined \footnote{Referring to the notation in Ref. \cite{Leask1986}, this is the interaction between the atom on ``sublattice 1'' with itself in the next unit cell}, and this interaction is small compared to the other exchange interactions. This suggests that the magnetic interactions in \dersio\ could be quasi-2D. Indeed, the layered structure of Er$^{3+}$ in \dersio\ also suggests that the magnetic interactions may be quasi-2D.

\section{Conclusions}
In this work we have used AC susceptibility, neutron diffraction, and inelastic neutron scattering to study the Ising-like antiferromagnetic order and field-induced behavior of the rare earth pyrosilicate, \dersio. AC susceptibility measurements with a field transverse to the Ising direction show a transition at a magnetic field strength of 2.65 T. Using neutron diffraction we have determined the magnetic structure of \dersio, which consists of moments pointing along a local direction which is close to the Ising direction determined by Leask \textit{et. al} \cite{Leask1986}. Our powder INS measurements reveal gapped excitations, one of which softens under an applied field and become gapless near 1 T. Due to both neutron scattering experiments being performed on powder samples, it is difficult to directly connect the transition observed in AC susceptibility on a single crystal to any signature in the neutron scattering data. However, we can state that there is a transition for a field applied perpendicular to the Ising axes, the zero-field INS spectra is consistent with Ising-like exchange, and the magnetic structure has been determined via neutron diffraction. This work sets the stage for future studies confirming the Ising-like nature of interactions in \dersio\ and for further study of the transverse field transition. Future work should include INS measurements on single crystals \cite{Nair2019} in order to further elucidate the nature of the transverse-field-induced transitions in this material.

\section{Acknowledgements}
This research was supported by the National Science Foundation Agreement No. DMR-1611217. The authors would like to acknowledge the assistance of Aaron Glock and Antony Sikorski in the sample synthesis. The authors acknowledge the Central Instrument Facility at Colorado State University for instrument access and training. The authors would also like to thank Gang Cao for enabling our use of millikelvin range AC susceptometry. The authors would also like to thank John McArthur of Quantum Design Japan for performing the DC magnetometry measurements presented in the appendix. Access to MACS was provided by the Center for High Resolution Neutron Scattering, a partnership between the National Institute of Standards and Technology and the National Science Foundation under Agreement No. DMR-1508249. A portion of this work used resources at the High Flux Isotope Reactor, a DOE Office of Science User Facility operated by Oak Ridge National Laboratory.

\clearpage
\section{Appendix}
\subsection{Heat Capacity Measurements}
\label{Cp_SI}

\begin{figure}[htp]
\centering
\includegraphics[width=0.9\columnwidth]{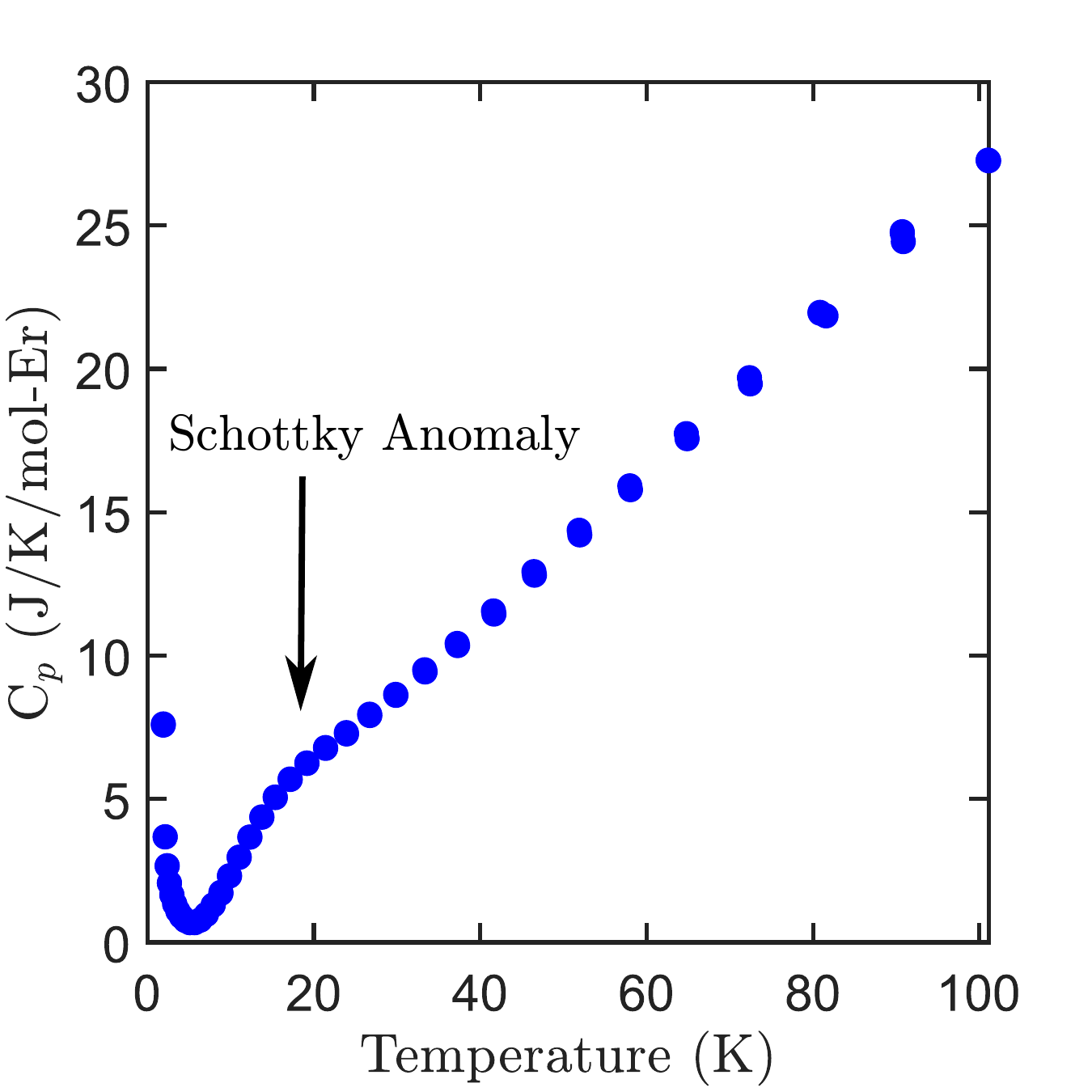}
\caption{Specific heat measured on a powder sample of \dersio. } \label{fig_cp}
\end{figure}

The heat capacity for 1.8 K to 100 K is shown in Fig.~\ref{fig_cp}. A broad feature on top of the phonon contribution is observed. The peak of this feature occurs at $\sim$ 16 K and can be attributed to a crystal field Schottky anomaly around 39 K, consistent with the lowest crystal electric field level measured by Leask \textit{et. al.} at 27 cm$^{-1}$ (39 K) \cite{Leask1986}.

\subsection{Magnetic Structure Refinement}
\label{MagStruc_SI}
\begin{figure}[htp]
\centering
\includegraphics[width=0.9\columnwidth]{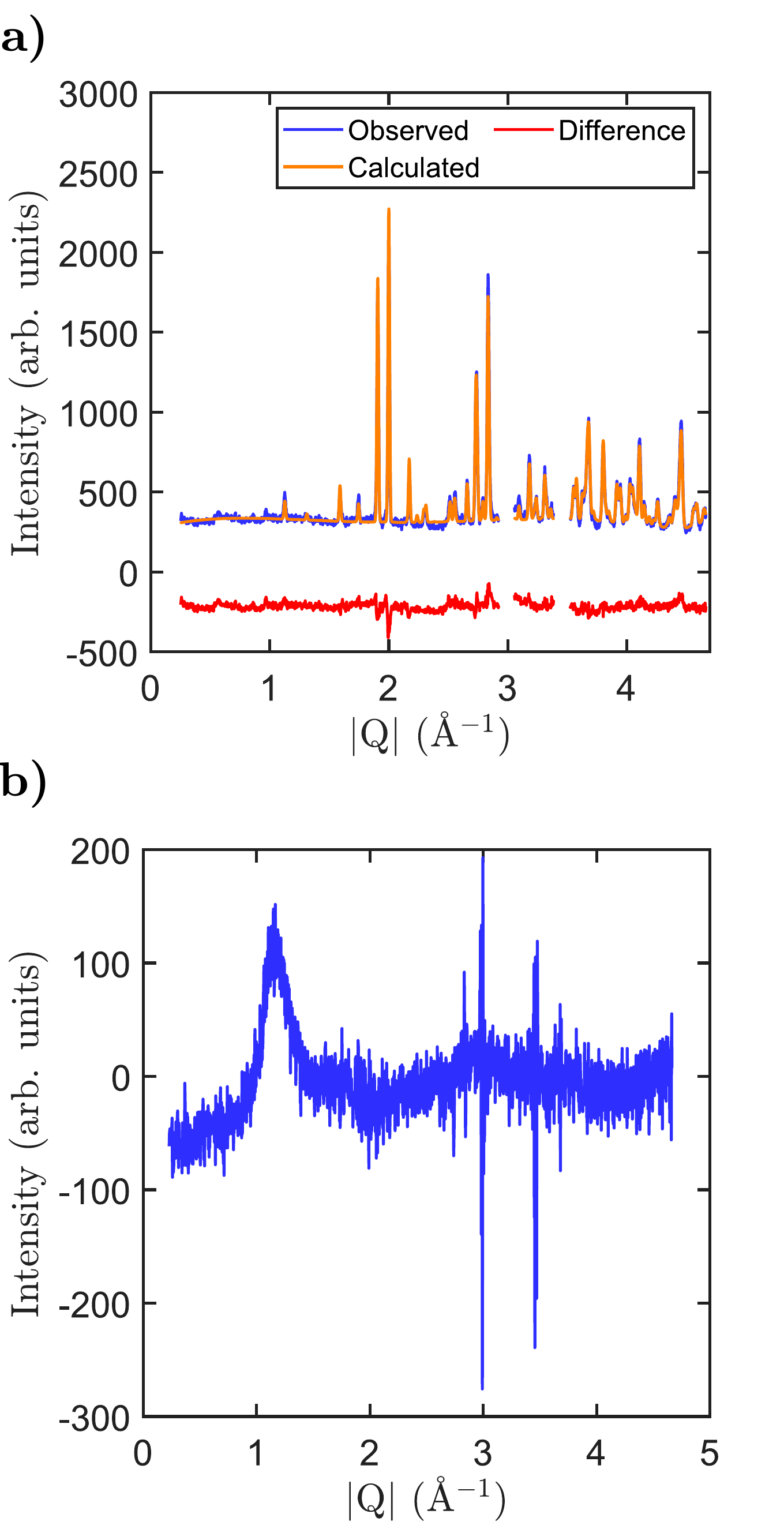}
\caption{a) Neutron diffraction data obtained using HB-2A at 10 K. The fit was performed after removing peaks from the Cu canister. b) Neutron diffraction data obtained using HB-2A with the 10 K data subtracted from the 2 K data showing the diffuse scattering expected in the vicinity of a second order phase transition, with a peak centered at ~1.2 \AA .}\label{fig_2K_10KCombine}
\end{figure}
Additional neutron diffraction data obtained on a powder sample of \dersio\ is shown in Fig.~\ref{fig_2K_10KCombine}. The data at 10 K (Fig.~\ref{fig_2K_10KCombine}a) was used to perform a Rietveld refinement on the nuclear structure of \dersio\ with the Cu peaks masked. This yielded the following lattice parameters at 10K for \dersio\ : $a$ = 4.6808(8) \AA, $b$ = 5.5566(2) \AA, $c$ = 10.7864(4) \AA, $\alpha$ = 90$^{\circ}$, $\beta$ = 90$^{\circ}$, $\gamma$ = 96.064(2)$^{\circ}$. These values found for the nuclear structure of \dersio\ were used in the refinement of the magnetic structure in Fig.~\ref{fig_elastic}a. The neutron diffraction data shown in Fig.~\ref{fig_2K_10KCombine}b is a subtraction of the 10 K data from the 2 K data. This shows the development of diffuse scattering expected in the vicinity of a second order phase transition. This is consistent with the magnetic ordering transition observed in \dersio\ at 1.9 K.

\begin{table*}[htp]
 \centering
 \begin{tabular}{|c c c c | c c c | c c c | c c c | c |} 
 \hline
 \multicolumn{4}{|c|}{Atom Info} & \multicolumn{3}{|c|}{$\psi_{10}$} & \multicolumn{3}{|c|}{$\psi_{11}$} & \multicolumn{3}{|c|}{$\psi_{12}$} & Moment Direction  \\
 \hline
 Atom & x & y & z & m$_{a}$ & m$_{b}$ & m$_{c}$ & m$_{a}$ & m$_{b}$ & m$_{c}$ & m$_{a}$ & m$_{b}$ & m$_{c}$ & (\textbf{a}, \textbf{b}, \textbf{c} basis) \\ [1.0ex] 
 \hline
 1 & 0.88829 & 0.09318 & 0.34934 & 1 & 0 & 0 & 0 & 1 & 0 & 0 & 0 & 1 & (5.72(3), -2.34(6), -1.45(6))\\ 
 \hline
 2 & 0.11171 & 0.40682 & 0.84934 & 1 & 0 & 0 & 0 & 1 & 0 & 0 & 0 & -1 & (5.72(3), -2.34(6), 1.45(6))\\
 \hline
 3 & 0.11171 & 0.90682 & 0.65066 & -1 & 0 & 0 & 0 & -1 & 0 & 0 & 0 & -1 & (-5.72(3), 2.34(6), 1.45(6)) \\
 \hline
 4 & 0.88829 & 0.59318 & 0.15066 & -1 & 0 & 0 & 0 & -1 & 0 & 0 & 0 & 1 & (-5.72(3), 2.34(6), -1.45(6))\\
 \hline
   &  &  &  & \multicolumn{3}{|c|}{C$_{10}$ = 5.72(3)} &\multicolumn{3}{|c|}{C$_{11}$ = -2.34(6)} & \multicolumn{3}{|c|}{C$_{12}$ = -1.45(6)} & \\
 \hline
\end{tabular}
\caption{A table showing the positions (x, y, z) of the four Er$^{3+}$ sublattices in the unit cell, the representation for each basis vector in terms of the (non-orthogonal) crystallographic \textit{a},\textit{b}, \textit{c} axes, the refined contributions (C$_{10}$, C$_{11}$, C$_{12}$) of each basis vector, and the resulting refined moment vectors for each Er$^{3+}$ site. } \label{table_bv}
\end{table*}

The basis vector composition of the three basis vectors ($\psi_{10}$, $\psi_{11}$, $\psi_{12}$) in the $\Gamma_{4}$ irreducible representation are shown in Table~\ref{table_bv}. These basis vectors are part of the $\Gamma_{4}$ irreducible representation which relates to the "m" point group symmetry and magnetic space group P2$_{1}$'/c.

\subsection{AC Susceptibility and Magnetometry}

\subsubsection{Measurements along the transverse direction}
\label{AC_SI}

\begin{figure*}[!t]
\includegraphics[width=1.9\columnwidth]{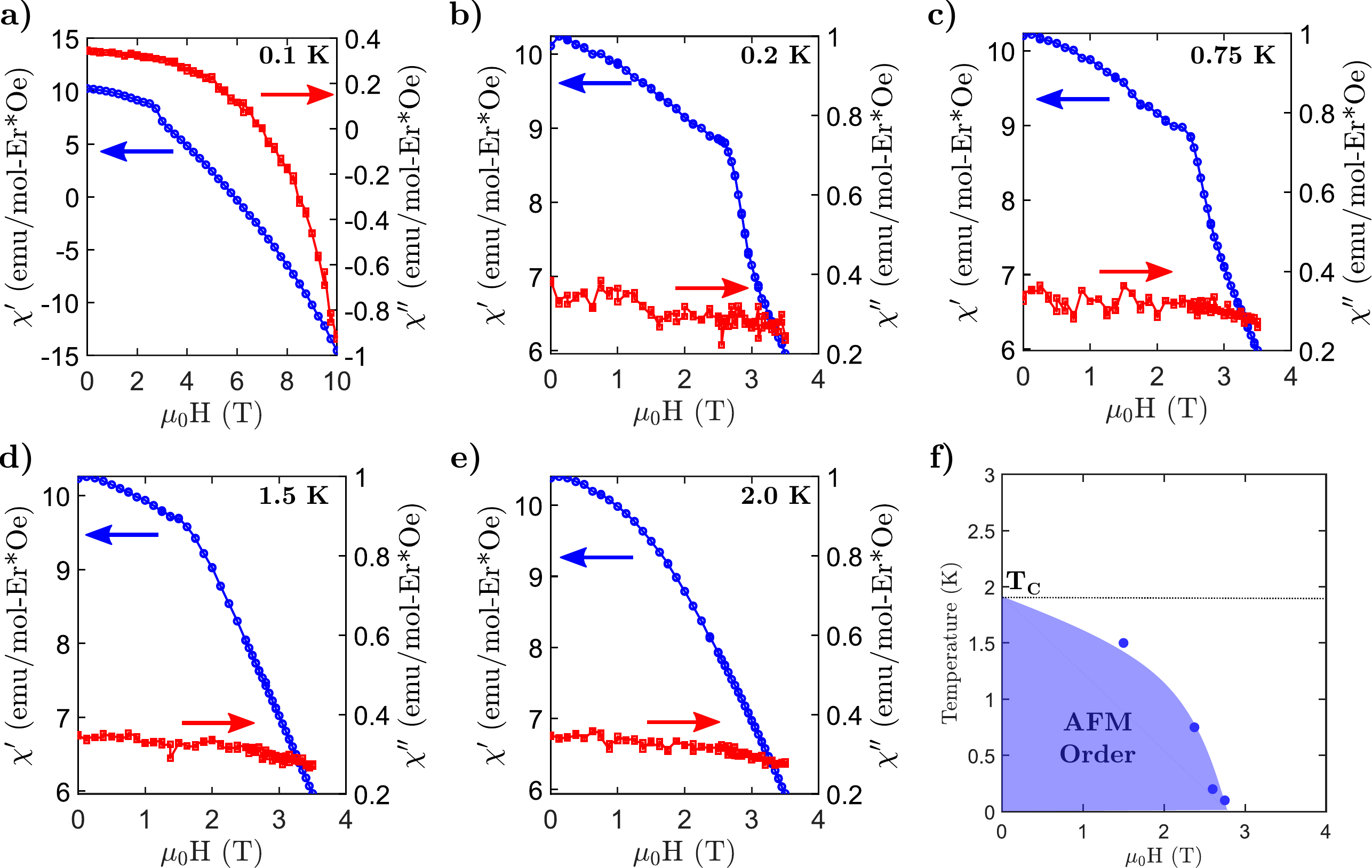}
\caption{a)-e) AC susceptibility data measured at several temperatures with $f = 1000$ Hz, $H_{AC}$ = 2 mT up to a DC field of 10 T (panel a) or 3.5 T (panels b through e) with the field applied perpendicular to the Ising direction (i.e. applied along the $x$ direction).  The field-induced transitions are observable as a kink in the real component of the susceptibility ($\chi^{\prime}$).  f) Phase diagram for \dersio\ based on the AC susceptibility measurements shown in panels a-e. The transition points represent the points where the real part of the susceptibility ($\chi^{\prime}$) abruptly changes due to the transition. The shaded region indicates AFM order. } \label{fig_extraACdata}
\end{figure*}

Additional AC susceptibility measurements for a field applied perpendicular to the Ising direction are shown in and Fig.~\ref{fig_extraACdata}. All data in this shown in this section is measured with $f = 1000$ Hz and $H_{AC}$ = 2 mT. The data shown in Fig.~\ref{fig_extraACdata}a demonstrates the susceptibility up to a DC field of 10 T. The transition observed in Fig.~\ref{fig_magnetometry_dersio} at 2.65 T is also observed here. At high fields the signal becomes diamagnetic, partly from the sample, partly from the mount and glue, and is likely due to the small moment along the transverse field direction which becomes saturated, leading to a small paramagnetic response. The data shown in Fig.~\ref{fig_extraACdata}b-e consists of four constant-temperature field sweeps (increasing field) and demonstrates how the 2.65 T transverse field transition evolves with temperature. There is little change in the imaginary component of the susceptibility ($\chi^{\prime\prime}$) as the temperature increases, but the real component of the susceptibility ($\chi^{\prime}$) changes, with the transition moving down in field as the temperature is increased. Measurements with the field decreasing were not performed for these temperatures and therefore no statement regarding the hysteresis observed in Fig.~\ref{fig_magnetometry_dersio} can be made.

\subsubsection{Measurements along (4$\overline{\textbf{9}}$1)}
\begin{figure}[htp]
\centering
\includegraphics[width=0.85\columnwidth]{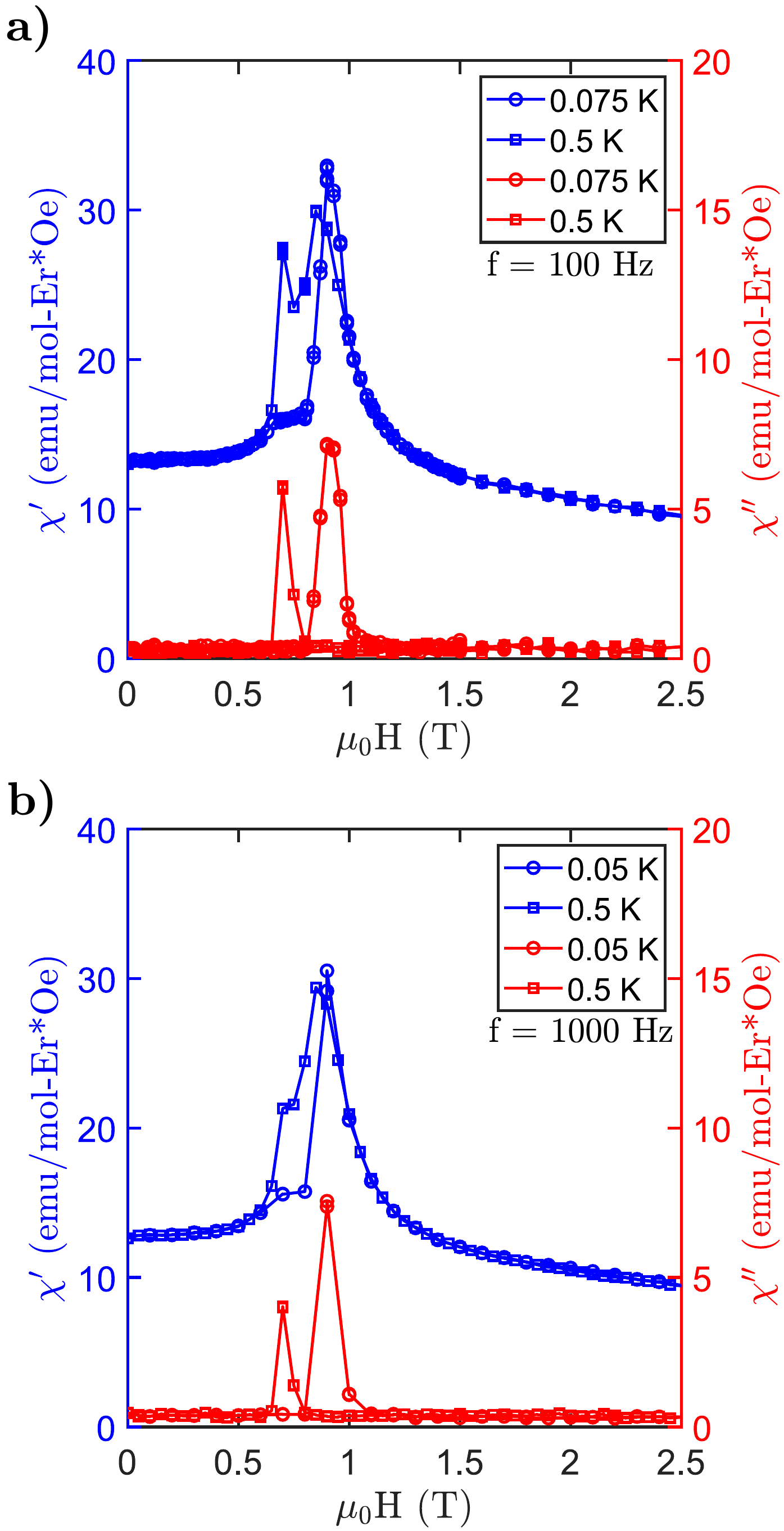}
\caption{a) AC susceptibility at 0.075 K and 0.5 K performed with a frequency of 100 Hz in an DC magnetic field applied 37$^{\circ}$ from the average Ising direction, approximately along the [4$\overline{9}$1] direction. and a 2 Oe AC magnetic field. b) AC susceptibility at 0.05 K and 0.5 K performed with a frequency of 1000 Hz in an DC magnetic field applied 37$^{\circ}$ from the average Ising direction, approximately along the [4$\overline{9}$1] direction. and a 2 mT AC magnetic field.}\label{fig_ACSuscept_SI}
\end{figure}

\begin{figure}[t]
\centering
\includegraphics[width=\columnwidth]{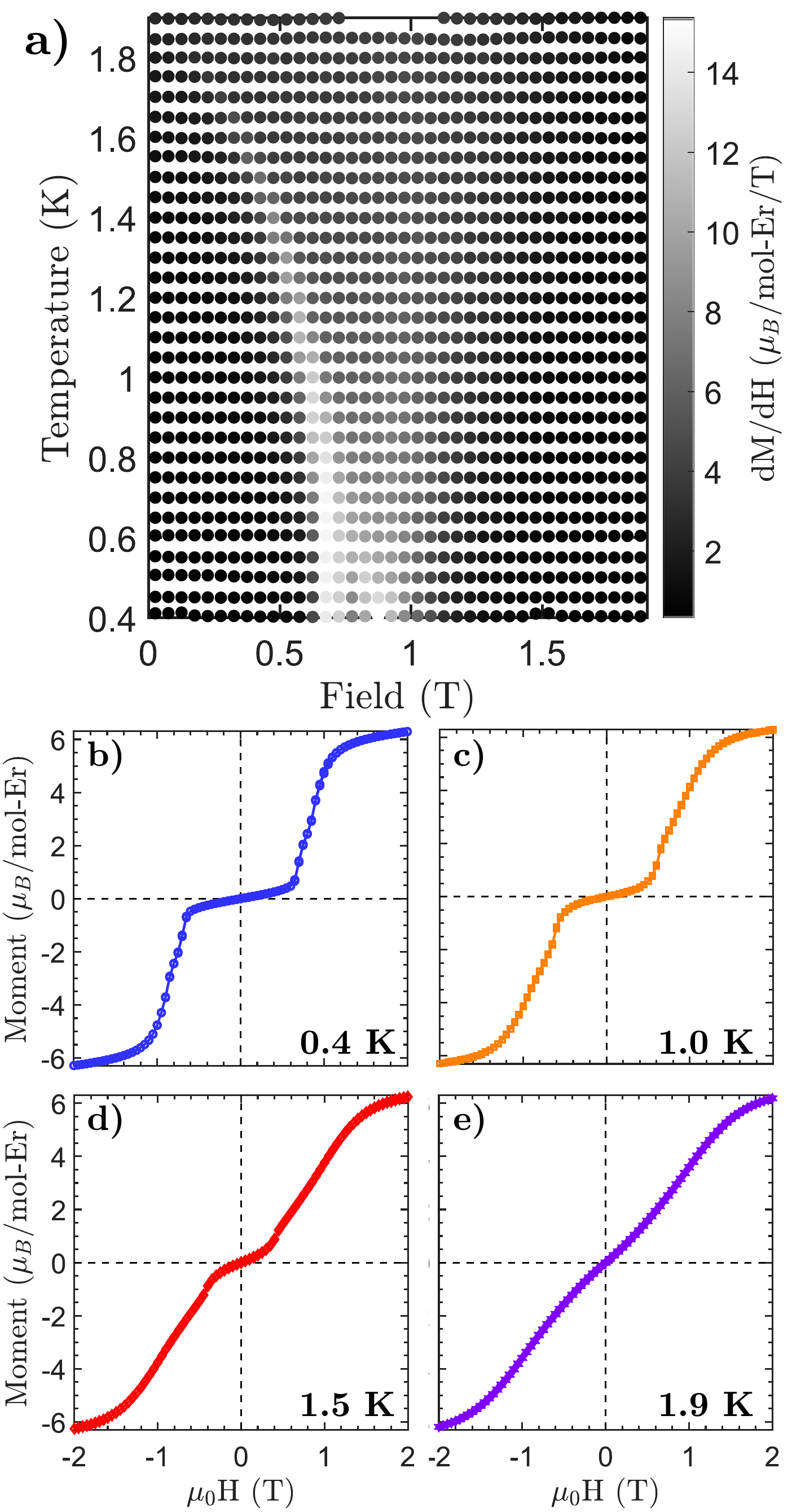}
\caption{a) A grayscale view of numerous constant-temperature field sweeps for the field-derivative of the DC magnetic susceptibility, showing a transition at 0.65~T that decreases in field as the temperature is increased. b-e) Characteristic field sweeps from the data shown in panel a). } \label{fig_dcmagnetometry}
\end{figure}

As mentioned in the main text, a previous version of this manuscript showed AC susceptibility data that was presented as a field applied transverse to the Ising direction, but it was later discovered that the measurements were actually for a field applied 37$^{\circ}$ from the average Ising direction, approximately along the (4$\overline{9}$1) direction. We have included these data here (Fig.~\ref{fig_ACSuscept_SI}). A comparison of data with $f = 100$ Hz at $T = 0.075$ K (originally shown in the main text) and 0.5 K is shown in Fig.~\ref{fig_ACSuscept_SI}a. The data at 0.075 K shows a sharp peak in both the real ($\chi$') and imaginary ($\chi^{\prime\prime}$) components at 0.9 T, with a small shoulder near 0.65 T. This shoulder has no significant imaginary component. This changes at 0.5 K, where the shoulder increase significantly in magnitude in the real part of the susceptibility. We also see that the peak in the imaginary component shifts to coincide with the shoulder at 0.65 T. Similar changes are observed for a comparison of data with $f = 1000$ Hz and $T = 0.05$ K and 0.5 K, shown in Fig.~\ref{fig_ACSuscept_SI}b. Overall, there is not a large difference between data measured at $f = 100$ Hz and $f = 1000$ Hz. One notable signature is that the transition at 0.65 T changes in magnitude with higher frequencies, indicating there may be some frequency dependence to this transition. The cause for these changes is not currently known, but may indicate a more complex field-induced phase diagram.

The DC magnetometry measurements that enabled us to identify the issue with the alignment are shown in Fig.~\ref{fig_dcmagnetometry}. These measurements were performed with a $\prescript{3}{}{\textnormal{He}}$ insert for the Quantum Design MPMS3. Measurements were performed for a field applied 37$^{\circ}$ from the average Ising direction, approximately along the [4$\overline{9}$1] direction. The fact that the measured moment is so large along this field direction (6 $\mu_{B}$) led to the discovery that the sample was misaligned, as the transverse field direction would be expected to have a maximum moment of ~1.3 $\mu_{B}$ (based off Leask \textit{et. al}'s g-tensor value of 2.6 for the transverse field direction).


\clearpage
\subsection{MACS Data}
\label{MACS_SI}

\begin{figure}[htp]
\centering
\includegraphics[width=0.8\columnwidth]{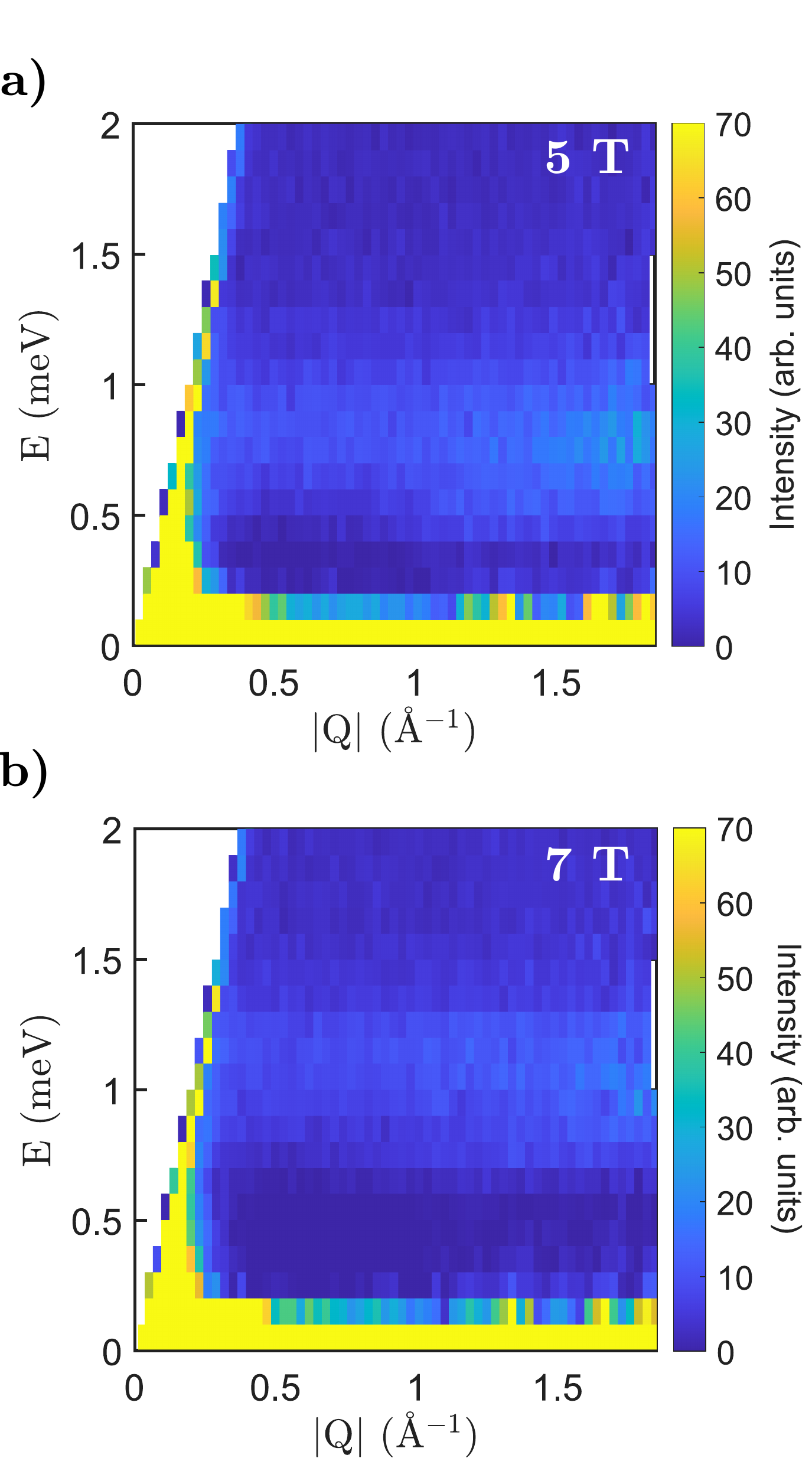}
\caption{a) Energy vs. |Q| slices (T$_{avg}$ = 0.16 K) at $\mu_{0}$H = 5 T. Branch 2 has increased relative to 3 T (Fig.~\ref{fig_inelastic}e) due to Zeeman splitting. b) Energy vs. |Q| slices (T$_{avg}$ = 0.16 K) at $\mu_{0}$H = 7 T. Branch 2 has increased relative to 5 T, causing the energy window displayed to be increased from 1 meV to 2 meV.}\label{fig_5T_7TCombine}
\end{figure}

Additional inelastic slices from the MACS data at 5 T (panel a) and 7 T (panel b) are shown in Fig.~\ref{fig_5T_7TCombine}. Both sets of data show Branch 2 increasing in energy relative to the 3 T data shown in Fig.~\ref{fig_inelastic}e. Note, in Fig.~\ref{fig_5T_7TCombine} the energy window was increased from 1 meV to 2 meV to allow all of Branch 2 to be visible.

\clearpage
%
%


\clearpage
%
%

%
%

\end{document}